\documentclass[%
 reprint,
superscriptaddress,
showkeys,
 amsmath,amssymb,
 aps,
 prl
]{revtex4-2}

\usepackage{graphicx}
\usepackage{dcolumn}
\usepackage{bm}
\usepackage[colorlinks=true,urlcolor=magenta,linkcolor=magenta,citecolor=cyan]{hyperref}
\usepackage{verbatim}
\usepackage[utf8]{inputenc}
\usepackage{layouts}
\usepackage{multirow}
\usepackage[dvipsnames]{xcolor}
\usepackage{hhline} 
\usepackage[separate-uncertainty=true]{siunitx}
\usepackage[version=3]{mhchem} 
\usepackage{wrapfig}
\usepackage{braket}
\usepackage{layouts}
\usepackage{float}
\usepackage{leftidx} 
\usepackage{amsmath,mathtools}
\usepackage{amssymb}
\usepackage{chngcntr}

\begin{document}


\title[\texttt{achemso} demonstration]
{Nonlinear Dispersion Relation and Out-of-Plane Second Harmonic Generation in\\ MoSSe and WSSe Janus Monolayers}


\author{Marko M. Petri\'{c}}
\email{Marko.Petric@wsi.tum.de}
\affiliation{Walter Schottky Institut and Department of Electrical and Computer Engineering, Technische Universit\"{a}t M\"{u}nchen, Am Coulombwall 4, 85748 Garching, Germany}
\affiliation{Munich Center for Quantum Science and Technology (MCQST), Schellingstrasse 4, 80799 Munich, Germany}

\author{Viviana Villafañe}
\affiliation{Walter Schottky Institut and Physik-Department, Technische Universit\"{a}t M\"{u}nchen, Am Coulombwall 4, 85748 Garching, Germany}
\affiliation{Munich Center for Quantum Science and Technology (MCQST), Schellingstrasse 4, 80799 Munich, Germany}

\author{Paul Herrmann}
\affiliation{Institute of Solid State Physics, Friedrich Schiller University Jena, Max-Wien-Platz 1, 07743 Jena, Germany}

\author{Amine Ben Mhenni}
\affiliation{Walter Schottky Institut and Physik-Department, Technische Universit\"{a}t M\"{u}nchen, Am Coulombwall 4, 85748 Garching, Germany}
\affiliation{Munich Center for Quantum Science and Technology (MCQST), Schellingstrasse 4, 80799 Munich, Germany}

\author{Ying Qin}
\affiliation{Materials Science and Engineering, School for Engineering of Matter, Transport and Energy, Arizona State University, Tempe, Arizona 85287, USA}

\author{Yasir Sayyad}
\affiliation{Materials Science and Engineering, School for Engineering of Matter, Transport and Energy, Arizona State University, Tempe, Arizona 85287, USA}

\author{Yuxia Shen}
\affiliation{Materials Science and Engineering, School for Engineering of Matter, Transport and Energy, Arizona State University, Tempe, Arizona 85287, USA}

\author{Sefaattin Tongay}
\email{sefaattin.tongay@asu.edu}
\affiliation{Materials Science and Engineering, School for Engineering of Matter, Transport and Energy, Arizona State University, Tempe, Arizona 85287, USA}

\author{Kai M\"{u}ller}
\affiliation{Walter Schottky Institut and Department of Electrical and Computer Engineering, Technische Universit\"{a}t M\"{u}nchen, Am Coulombwall 4, 85748 Garching, Germany}
\affiliation{Munich Center for Quantum Science and Technology (MCQST), Schellingstrasse 4, 80799 Munich, Germany}

\author{Giancarlo Soavi}
\email{giancarlo.soavi@uni-jena.de}
\affiliation{Institute of Solid State Physics, Friedrich Schiller University Jena, Max-Wien-Platz 1, 07743 Jena, Germany}
\affiliation{Abbe Center of Photonics, Friedrich Schiller University Jena, Albert-Einstein-Straße 6, 07745 Jena, Germany\\}

\author{Jonathan J. Finley}
\affiliation{Walter Schottky Institut and Physik-Department, Technische Universit\"{a}t M\"{u}nchen, Am Coulombwall 4, 85748 Garching, Germany}
\affiliation{Munich Center for Quantum Science and Technology (MCQST), Schellingstrasse 4, 80799 Munich, Germany}

\author{Matteo Barbone}
\email{Matteo.Barbone@wsi.tum.de}
\affiliation{Walter Schottky Institut and Department of Electrical and Computer Engineering, Technische Universit\"{a}t M\"{u}nchen, Am Coulombwall 4, 85748 Garching, Germany}
\affiliation{Munich Center for Quantum Science and Technology (MCQST), Schellingstrasse 4, 80799 Munich, Germany}

\begin{abstract}
Janus transition metal dichalcogenides are an emerging class of atomically thin materials with engineered broken mirror symmetry that gives rise to long-lived dipolar excitons, Rashba splitting, and topologically protected solitons. They hold great promise as a versatile nonlinear optical platform due to their broadband harmonic generation tunability, ease of integration on photonic structures, and nonlinearities beyond the basal crystal plane. Here, we study second and third harmonic generation in MoSSe and WSSe Janus monolayers. We use polarization-resolved spectroscopy to map the full second-order susceptibility tensor of MoSSe, including its out-of-plane components. In addition, we measure the effective third-order susceptibility, and the second-order nonlinear dispersion close to exciton resonances for both MoSSe and WSSe at room and cryogenic temperatures. Our work sets a bedrock for understanding the nonlinear optical properties of Janus transition metal dichalcogenides and probing their use in the next-generation on-chip multifaceted photonic devices. 
\end{abstract}
\maketitle

Nonlinear optics with atomically thin and layered materials has proven to be a powerful spectroscopic approach to study the physics of excitons confined at the ultimate thickness limit, as well as a fruitful playground to test novel ultrafast and ultrathin optical devices \cite{dogadov2022}. Parametric harmonic generation has been used to probe interlayer excitons in homo-\cite{shree2021} and hetero-bilayers \cite{paradisanos2022}, and to detect quantum interference pathways and strong coupling in monolayer transition metal dichalcogenides (TMDs) \cite{lin2019}. Furthermore, nonlinear optical properties of layered materials are adding a rapidly growing number of applications in optics, which already include electrical \cite{soavi2018} and all-optical \cite{klimmer2021} ultrafast and broadband frequency converters, miniaturized logic gates \cite{zhang2022a, li2022}, photonic integrated gas sensors \cite{an2020}, nonlinear holograms \cite{dasgupta2019}, and ultrathin quantum sources based on spontaneous parametric down conversion (SPDC) \cite{guo2023}.

Lately, a new class has expanded the family of atomically thin materials: Janus transition metal dichalcogenide (Janus TMD) monolayers are van der Waals materials with different chalcogen atoms within the crystal unit cell, which breaks mirror symmetry, and creates an out-of-plane electrical polarity due to charge imbalance \cite{li2017, wang2018, shi2018}. The structural mirror-symmetry breaking in Janus TMD monolayers gives rise to physical effects and functionalities beyond those offered by conventional TMDs, such as Rashba spin-splitting \cite{cheng2013, cheng2016, yao2017, hu2018, adhibulilabsor2018, patel2022}, and topologically-protected and localized solitonic spin textures \cite{yuan2020, liang2020, cui2020}. Moreover, due to the reduced overlap between electron and hole wavefunctions, Janus TMD monolayers host longer-lived dipolar excitons than mirror-symmetric TMDs \cite{jin2018, li2019, long2019, zheng2021}, making them a potential platform for exciton Bose-Einstein condensation and other strongly correlated many-body states \cite{guo2022}. Since their theoretical prediction in 2013 \cite{cheng2013} and first synthesis in 2017 \cite{lu2017, zhang2017}, a multitude of morphotaxy-based techniques \cite{lam2022} have emerged breaching bottlenecks toward the synthesis of high-quality MoSSe and WSSe Janus monolayers \cite{lin2020, trivedi2020, guo2021, jang2022, qin2022, gan2022, harris2023}. Further studies have already provided insight into their vibrational properties \cite{petric2021, zhang2021}, interlayer coupling strength \cite{zhang2020a, zheng2021, zheng2022}, exciton complexes \cite{feuer2023}, second harmonic generation \cite{bian2022}, and dipole emission profile \cite{zhang2022}. 

Atomically thin crystals provide two distinct advantages compared to conventional bulk materials used in nonlinear optics. First, a high degree of compatibility with on-chip photonic architectures \cite{youngblood2016, tonndorf2017, guo2022a, ngo2022}. Second, their atomically thin nature combined with a strong nonlinear response \cite{autere2018, dogadov2022} allows for efficient frequency conversion without phase matching constraints \cite{zhao2016, trovatello2021}, and thus, provides a viable route towards ultra-broadband classical and quantum sources of light. Janus TMD monolayers further expand such advantages by respectively adding out-of-plane components in the nonlinear optical response \cite{wei2019, strasser2022, pike2022}, which introduce an additional functionality to be harnessed in vertical photonics structures \cite{kleemann2017, stuhrenberg2018}, and widening the spectral region of resonantly enhanced nonlinear optics to different areas of the visible and near-infrared light \cite{zhang2020b}. To capitalize on such advantages, understanding the wavelength-dependent second harmonic generation (SHG) efficiency (i.e., the nonlinear dispersion) and mapping the corresponding susceptibility tensor in Janus TMD monolayers is essential.

In this work, we address both these open issues and provide a detailed study of the out-of-plane second-order nonlinear susceptibility in MoSSe Janus monolayer, using a simplified approach that fully maps the susceptibility tensor using a high numerical aperture (NA) objective. Furthermore, we measure third harmonic generation (THG) and the wavelength-dependent SHG efficiency of MoSSe and WSSe Janus monolayers, revealing the $A$ and $B$ exciton resonances at room and cryogenic temperatures.

\section*{Results and Discussion}

\begin{figure*}[ht!]
\centering
\includegraphics{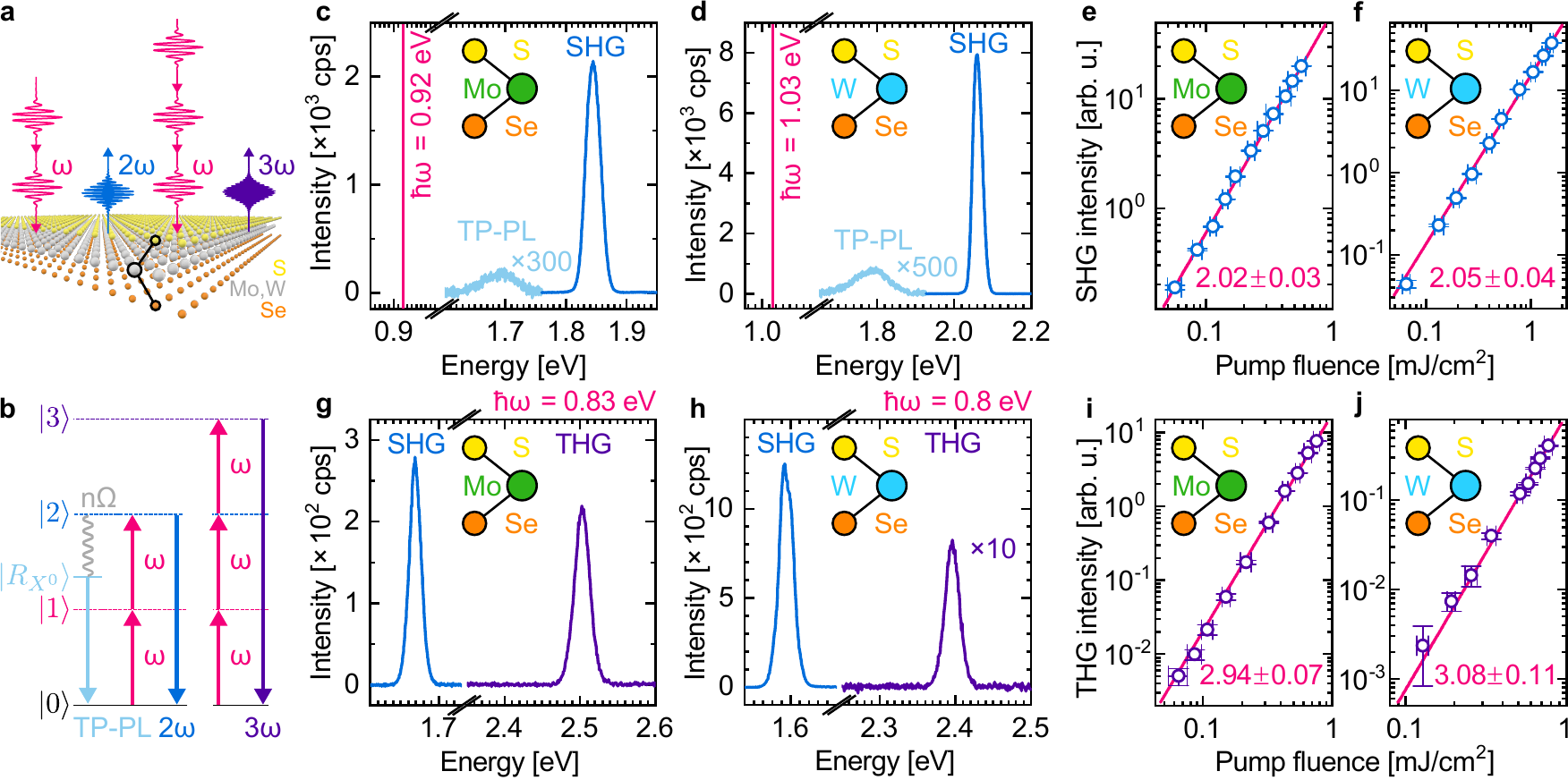}
\caption{\textbf{SHG and THG in MoSSe and WSSe Janus monolayers.} \textbf{(a)} Schematic representation of a Janus TMD monolayer and the reflection measurement geometry. \textbf{(b)} Schematic representation of SHG (blue), THG (purple), and TP-PL (light blue) transitions under two- and three-photon excitation with angular frequency $\omega$ (magenta) between the excited levels $\ket{1}$, $\ket{2}$, $\ket{3}$, the real exciton level $\ket{R_{X^0}}$, linked via single- or multi-phonon channel $n\Omega$ to $\ket{2}$ (gray), and the ground state $\ket{0}$. \textbf{(c,d)} Measured spectra showing SHG (blue) and scaled TP-PL (light blue) peaks stemming from \textbf{(c)} MoSSe and \textbf{(d)} WSSe Janus monolayer upon two-photon excitation (magenta). \textbf{(e,f)} Respective SHG peak intensities as a function of pump fluence show quadratic increase as extracted from the power law fit (magenta). \textbf{(g,h)} THG (purple) and SHG (blue) spectra of \textbf{(g)} MoSSe Janus monolayer with excitation energy $\hbar\omega = \SI{0.83}{\electronvolt}$ and \textbf{(h)} WSSe Janus monolayer with excitation energy $\hbar\omega = \SI{0.8}{\electronvolt}$. \textbf{(i,j)} Respective THG peak intensities as a function of pump fluence show cubic slopes as extracted from the power law fit (magenta).}
\label{fig:Figure1}
\end{figure*}

Janus TMD monolayers were fabricated via a room temperature selective epitaxy atomic replacement method \cite{trivedi2020}. In this process, the top-layer selenium atoms in previously grown MoSe$_2$ (on Si/SiO$_{2}$) and WSe$_2$ (on Al$_{2}$O$_{3}$) parent monolayers are replaced by sulfur atoms, resulting in MoSSe and WSSe Janus monolayers, respectively, schematically represented in Figure \ref{fig:Figure1}a. We studied their nonlinear optical response at room and cryogenic temperatures by focusing linearly polarized pulsed light of angular frequency $\omega$ onto the crystal, and detecting the outgoing light in reflection geometry, as illustrated in Figure \ref{fig:Figure1}a. In particular, we used 100-fs pulses with a repetition rate of $\SI{80}{\mega\hertz}$ generated by a tunable optical parametric oscillator (OPO) and pumped by a mode-locked Ti:sapphire laser (see \hyperref[sec:Methods]{Methods}). 

MoSSe and WSSe Janus monolayers are non-centrosymmetric crystals \cite{cheng2013}, and thus, possess nonvanishing even-order terms in the polarization vector expansion. As a result, they exhibit a distinct second harmonic (SH) overtone induced by a light field of frequency $\omega$ \cite{shen2003, boyd2008a}:

\begin{equation}
P_{i}^{(2)}(2\omega) = \epsilon_0\sum_{j,k}\chi_{ijk}^{(2)}(2\omega;\omega,\omega)E_j(\omega)E_k(\omega)
\label{eq:SHGP}
\end{equation}

here, Cartesian components of the second-order polarization vector $\mathbf{P}^{(2)}(2\omega)$ capture the material's response to the interaction with an external optical field $\mathbf{E}(\omega)$. The two vectors are linked via a third-rank tensor $\boldsymbol\chi^{(2)}(2\omega;\omega,\omega)$, which can be directly deduced from the crystal symmetry. SHG is a parametric process where the annihilation of two photons of frequency $\omega$ is followed by the creation of a single photon of frequency $2\omega$. The photons link (virtual or real) excited states $\ket{1}$ and $\ket{2}$ to the electronic ground state  $\ket{0}$ \cite{franken1961, boyd2008a}, as illustrated in Figure \ref{fig:Figure1}b. In contrast, THG exists irrespective of the inversion symmetry and involves three annihilated photons of energy $\hbar\omega$ and a created photon of energy $3\hbar\omega$ from a state $\ket{3}$ denoted in purple in Figure \ref{fig:Figure1}b. Analogously to Eq. (\ref{eq:SHGP}), third-order polarization properties of a crystal are dictated by a fourth-rank tensor $\boldsymbol\chi^{(3)}(3\omega;\omega,\omega,\omega)$ \cite{boyd2008a}. In the case of TMDs, upon a two-photon excitation, the system may in addition relax via $n =1,2,3,...$ phonons $\Omega$ \cite{paradisanos2021} through a two-photon photoluminescence (TP-PL) involving a real exciton state $\ket{R_{X^0}}$ \cite{wang2015}, denoted in light blue.

Figure \ref{fig:Figure1}c (\ref{fig:Figure1}d) shows the detected light from MoSSe (WSSe) Janus monolayer at room temperature using $\hbar\omega = \SI{0.92}{\electronvolt}$ ($\hbar\omega = \SI{1.03}{\electronvolt}$) excitation energy. We choose these specific energies to separate the SHG and TP-PL signals. For both materials, we detect a pronounced SHG signal (blue) and less efficient TP-PL signal (light blue) at a photon energy corresponding to the $1s$ neutral exciton $X^0$ at ${\approx}\SI{1.69}{\electronvolt}$ (${\approx}\SI{1.79}{\electronvolt}$) \cite{petric2021}. Figures \ref{fig:Figure1}e and \ref{fig:Figure1}f show the power dependence of the SH signal for MoSSe and WSSe Janus monolayers, respectively, confirming the SH nature of the signal at $2\hbar\omega$ via the canonical quadratic increase in the SH peak power as a function of pump fluence \cite{boyd2008a, malard2013, kumar2013}. In addition, we measure circular polarization selection rules consistent with SHG (see \hyperref[sec:SuppInfo]{SI} Section \hyperref[sec:SuppInfo]{S1}).


We also observe THG signals: Figures \ref{fig:Figure1}g and \ref{fig:Figure1}h show room temperature SHG and THG spectra of MoSSe and WSSe Janus monolayers for excitation energies $\hbar\omega = \SI{0.83}{\electronvolt}$ and $\hbar\omega = \SI{0.8}{\electronvolt}$, respectively. To confirm the third harmonic (TH) nature of the signal, we again perform power-dependent measurements which exhibit power law dependencies with near cubic behavior, as expected \cite{wang2014, woodward2016, saynatjoki2017, rosa2018}. The results are presented in Figure \ref{fig:Figure1}i for MoSSe, and Figure \ref{fig:Figure1}j for WSSe Janus monolayer, respectively. The intensity of the nonlinear response is normally expected to drop with increasing nonlinear order. However, for low pumping energies, TMDs exhibit a higher THG compared to the SHG signal due to nearly rotationally invariant bands only corrected by trigonal warping of the valence and conduction bands \cite{kormanyos2013, saynatjoki2017}. MoSSe Janus monolayers conform to this behavior, exhibiting comparable SHG and THG signals, as shown in Figure \ref{fig:Figure1}g. Here, THG does not dominate over SHG because the latter is close to resonance with the $A$ exciton. Furthermore, we derive the value of the effective third-order susceptibility (see \hyperref[sec:SuppInfo]{SI} Section \hyperref[sec:SuppInfo]{S2}) and obtain $\chi_{\mathrm{eff}}^{(3)}(\hbar\omega = \SI{0.8}{\electronvolt}) = 1.55\times10^5\,\SI{}{\pico\meter^2/\volt^2}$ which is close to reported values for other TMDs \cite{woodward2016, autere2018, wang2014, rosa2018, saynatjoki2017}. In terms of harmonic efficiency, defined as the ratio of THG peak power to incident power, we obtain ${\sim}3\times10^{-11}$ for an incident power of $\SI{20}{\milli\watt}$. Interestingly, in a WSSe Janus monolayer the SHG intensity dominates over the THG, in contrast to all measurements on TMDs reported to date. For WSSe Janus monolayer, the measured effective third-order susceptibility is $\chi_{\mathrm{eff}}^{(3)}(\hbar\omega = \SI{0.83}{\electronvolt}) = 1.52\times10^5\,\SI{}{\pico\meter^2/\volt^2}$, and the harmonic efficiency is ${\sim}10^{-11}$ for an incident power of $\SI{20}{\milli\watt}$.

Next, we shift our attention to the study of the out-of-plane second-order nonlinear response of MoSSe Janus monolayer. In contrast to conventional TMDs, which belong to the $D_{3h}$ point group, the lack of mirror symmetry puts Janus TMD monolayers in the $C_{3v}$ point group, which displays five independent nonzero components $\chi_{xzx}^{(2)} = \chi_{yzy}^{(2)}$, $\chi_{xxz}^{(2)} = \chi_{yyz}^{(2)}$, $\chi_{zxx}^{(2)} = \chi_{zyy}^{(2)}$, $\chi_{zzz}^{(2)}$, $\chi_{yyy}^{(2)} = -\chi_{yxx}^{(2)} = -\chi_{xxy}^{(2)} = -\chi_{xyx}^{(2)}$. Furthermore, the Kleinman's symmetry condition in SHG always allows to permute the last two indices \cite{boyd2008a}; thus, $\chi_{xzx}^{(2)} = \chi_{xxz}^{(2)}$. Finally, we explicitly write the susceptibility tensor $\boldsymbol\chi_{C_{3v}}^{(2)}(2\omega;\omega,\omega)$ in contracted notation \cite{boyd2008a} by rewriting Eq. (\ref{eq:SHGP}) in a matrix form: 

\begin{widetext}
\begin{equation}
\begin{pmatrix}
    P_{x}^{(2)}(2\omega)\\
    P_{y}^{(2)}(2\omega)\\
    P_{z}^{(2)}(2\omega)
\end{pmatrix} = 
\epsilon_0
\begin{pmatrix}
    0 & 0 & 0 & 0 & \chi_{xxz}^{(2)} & -\chi_{yyy}^{(2)} \\
    -\chi_{yyy}^{(2)} & \chi_{yyy}^{(2)} & 0 & \chi_{xxz}^{(2)} & 0 & 0 \\
    \chi_{zxx}^{(2)} & \chi_{zxx}^{(2)} & \chi_{zzz}^{(2)} & 0 & 0 & 0 
\end{pmatrix}
\begin{pmatrix}
    E_x(\omega)E_x(\omega)\\
    E_y(\omega)E_y(\omega)\\
    E_z(\omega)E_z(\omega)\\
    2E_y(\omega)E_z(\omega)\\
    2E_x(\omega)E_z(\omega)\\
    2E_x(\omega)E_y(\omega)
\end{pmatrix}
\label{eq:matrix}
\end{equation}
\end{widetext}

\begin{figure}[ht!]
\centering
\includegraphics{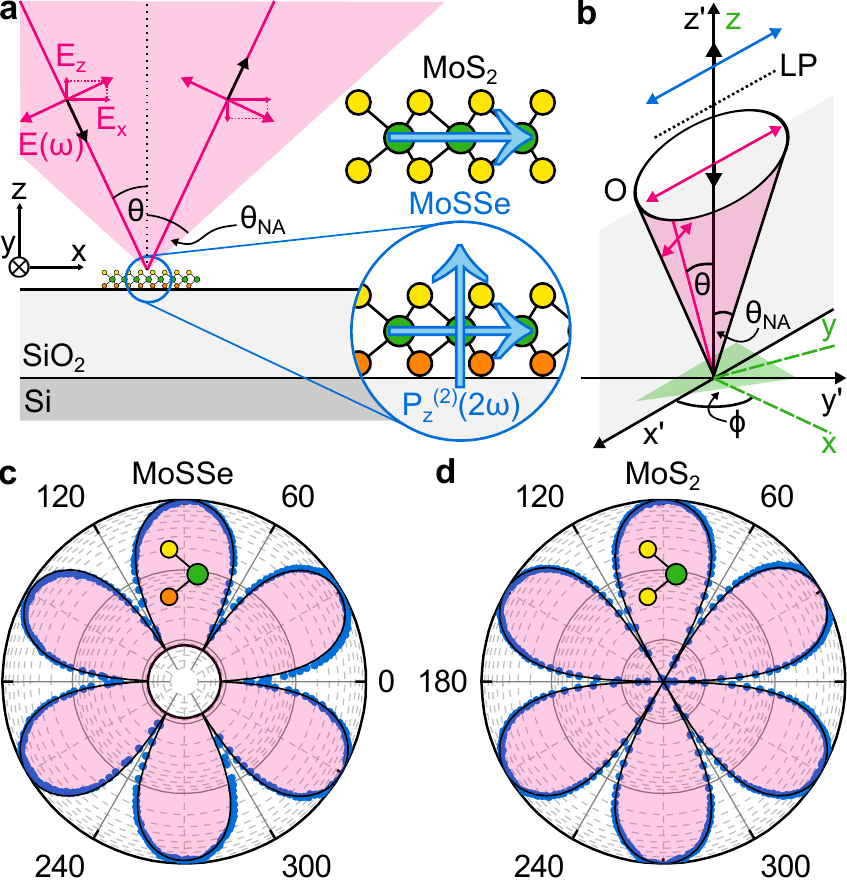}
\caption{\textbf{Out-of-plane polarization dipole in MoSSe Janus monolayer.} \textbf{(a)} Schematic representation of the out-of-plane SH excitation and detection. Magenta denotes light incident along an angle $\theta$ from the optical axis, limited to $\theta_{\mathrm{NA}}$ by the objective numerical aperture. MoSSe Janus monolayer possesses the $P_{z}^{(2)}(2\omega)$ component contrasting with the MoS$_2$ monolayer which only has an in-plane polarization component. \textbf{(b)} Schematic representation of the measurement configuration in a polar coordinate system. Excitation (magenta) and detection (blue) light have the same polarization entering and exiting the objective (O) determined with a linear polarizer (LP). Angle $\phi$ follows the crystal orientation. The crystal frame is highlighted in green. \textbf{(c,d)} Logarithmic polar plots of the normalized polarization-resolved SHG intensity of \textbf{(c)} Janus MoSSe and \textbf{(d)} MoS$_2$ monolayer. Measured intensity is represented with blue data points, the fitting curve in black, and the shaded region in pink is bound by the minimum of the fitting function.}
\label{fig:Figure2}
\end{figure}

Equation (\ref{eq:matrix}) implies two distinct features: (i) the presence of a nonvanishing out-of-plane $P_{z}^{(2)}(2\omega)$ polarization, as illustrated in the blue inset in Figure \ref{fig:Figure2}a, and (ii) the introduction of additional terms to the polarization vector components by an out-of-plane driving field $E_z(\omega)$, which can be activated by exciting the crystal at an angle, as illustrated in Figure \ref{fig:Figure2}a with a magenta line. 

The experimentally accessible SHG intensity components can be conveniently derived from the polarization $P_{x,y,z}^{(2)}(2\omega)$ by introducing a polar coordinate system, as shown in Figure \ref{fig:Figure2}b. The resulting SHG intensity components in the crystal frame of reference are: 

\begin{equation}
    I_x\propto\big|\chi_{xxz}^{(2)}\sin 2\theta + \chi_{yyy}^{(2)}\cos^2 \theta\sin 3\phi\big|^2
    \label{eq:Ix}
\end{equation}
\begin{equation}
    I_y\propto\big|\chi_{yyy}^{(2)}\cos^2 \theta\cos 3\phi\big|^2
    \label{eq:Iy}
\end{equation}
\begin{equation}
    I_z\propto\big|\chi_{zxx}^{(2)}\cos^2\theta + \chi_{zzz}^{(2)}\sin^2\theta\big|^2
    \label{eq:Iz}
\end{equation}

where $\theta$ is the incidence angle of excitation, and $\phi$ is the angle between the crystal zigzag direction ($x$-axis) and the $x'$ excitation and detection axis in the laser frame of reference \cite{pike2022}. In contrast, the SHG intensity components stemming from $D_{3h}$ TMDs are \cite{kumar2013, li2013, pike2022}:
\begin{equation}
I_x\propto\big|\chi_{yyy}^{(2)}\cos^2 \theta\sin 3\phi\big|^2
\label{eq:IxD3h}
\end{equation}
\begin{equation}
I_y\propto\big|\chi_{yyy}^{(2)}\cos^2 \theta\cos 3\phi\big|^2
\label{eq:IyD3h}
\end{equation}

To extract the susceptibility tensor components of Janus TMD monolayers, we measure the SHG intensity as a function of the rotation angle $\phi$ in a co-polarized configuration (see \hyperref[sec:Methods]{Methods}). This is equivalent to rotating the crystal in the laser frame of reference. We use the aforementioned equations to model the experimentally collected SHG light intensity $I_p = I_x\cos^2\theta + I_z\sin^2\theta$, which sums both components present in the plane of incidence. We used a high-NA objective (NA = 0.75), which enables large angles $\theta$ for strong out-of-plane driving and high detection efficiency of the out-of-plane polarization component. This approach simplifies the methods previously employed to retrieve non-planar components \cite{lu2017, xiao2018} and can easily be generalized for any material of a similar symmetry. Figure \ref{fig:Figure2}c plots normalized $I_p$ as a function of $\phi$ on a logarithmic scale measured from a MoSSe Janus monolayer excited at $\hbar\omega = \SI{0.885}{\electronvolt}$, away from the $A$ exciton resonance to avoid the TP-PL signal that could be unpolarized, and thus, could leave residual signals in the polarization-dependent SHG experiment that are difficult to deconvolve. Following Eqs. (\ref{eq:Ix}) and (\ref{eq:Iz}), we expect that out-of-plane susceptibility components $\chi_{xxz}^{(2)},\chi_{zxx}^{(2)}$, and $\chi_{zzz}^{(2)}$ contribute to the nonvanishing $I_p$ for all angles $\phi$. Indeed, the flower pattern 'opens up', never reaching zero, as shown in Figure \ref{fig:Figure2}c by the nonshaded region where the upper border represents the minimum value of the fitting function in black given by Eq. (\hyperref[sec:SuppInfo]{S33}) (see \hyperref[sec:SuppInfo]{SI} Section \hyperref[sec:SuppInfo]{S3}). Here, we take into account the angle-dependent substrate reflection and integrate the resulting intensity over the angle $\theta$ for the entire NA of the objective as explained in \hyperref[sec:SuppInfo]{SI} Section \hyperref[sec:SuppInfo]{S3}. In comparison, Figure \ref{fig:Figure2}d shows normalized polarization-resolved SHG intensity of a reference MoS$_2$ monolayer for the same excitation energy and the same-scaled logarithmic polar plot. $I_p$ follows the characteristic six-fold pattern given by Eq. (\ref{eq:IxD3h}) which goes to zero when excitation polarization is aligned with the zigzag crystal edge \cite{malard2013, kumar2013, li2013}, strongly contrasting with the MoSSe Janus monolayer in Figure \ref{fig:Figure2}c (see \hyperref[sec:SuppInfo]{SI} Section \hyperref[sec:SuppInfo]{S4}). 

\begin{figure*}[ht!]
\centering
\includegraphics{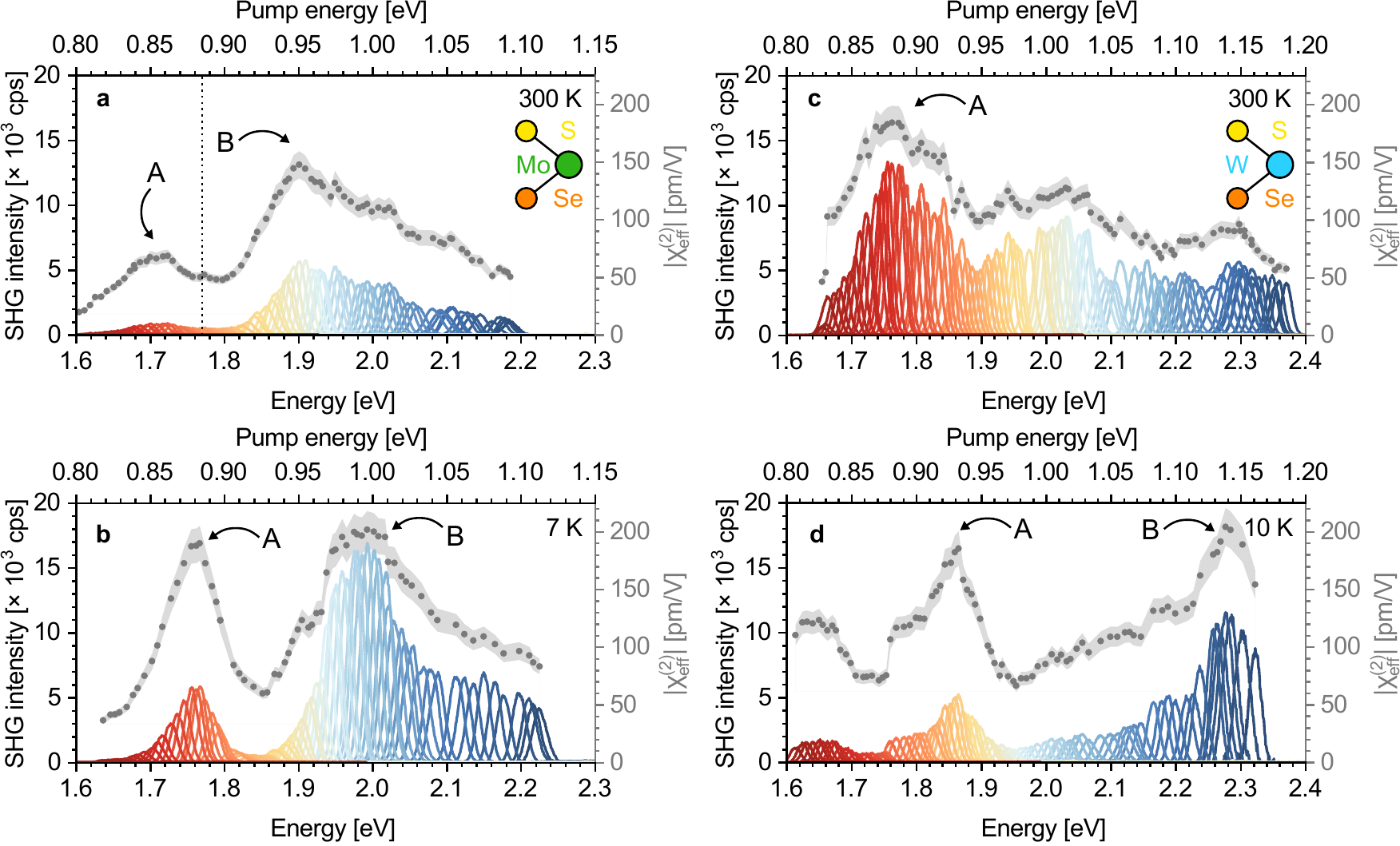}
\caption{\textbf{SHG enhancement at exciton resonances and $\chi^{(2)}_{yyy}$ dispersion.} \textbf{(a,b)} SHG spectra from MoSSe Janus monolayer as a function of two-photon energy (bottom $x$-axis) and pump energy (top $x$-axis) at \textbf{(a)} $T = \SI{300}{K}$ and \textbf{(b)} $T = \SI{7}{K}$ with $\chi^{(2)}_{yyy}$ dispersion represented in gray points within a light gray error band. \textbf{(c,d)} SHG spectra from WSSe Janus monolayer as a function of two-photon energy and pump energy at \textbf{(c)} $T = \SI{300}{K}$ and \textbf{(d)} $T = \SI{10}{K}$ with $\chi^{(2)}_{yyy}$ dispersion represented in gray points within a light gray error band. Assignments of $A$ and $B$ excitons are denoted by arrows.}
\label{fig:Figure3}
\end{figure*}

We now turn to disentangling the complex interplay of all susceptibility components contributing to the SHG intensity $I_p$. In order to give a quantitative estimate of the out-of-plane components compared to the in-plane components, we simplify the $\boldsymbol\chi^{(2)}$ tensor by imposing the Kleinman's symmetry assuming dispersionless response at both $\omega$ and $2\omega$ to obtain $\chi_{xxz}^{(2)} = \chi_{zxx}^{(2)}$ \cite{boyd2008a, lu2017}. This is possible since under these specific experimental conditions the SHG process is performed far off exciton resonances. We further reduce the number of parameters by introducing a $\chi_{zzz}^{(2)}/\chi_{xxz}^{(2)}$ ratio deduced from theoretical calculations \cite{pike2022}. As a final result, we obtain $\chi_{yyy}^{(2)}/\chi_{xxz}^{(2)} = 8.4$ at $\hbar\omega = \SI{0.885}{\electronvolt}$ pumping (see \hyperref[sec:SuppInfo]{SI} Section \hyperref[sec:SuppInfo]{S3}), which is comparable to theoretically calculated values \cite{pike2022}. Thus, with our simple approach we are able to fully map the second-order susceptibility tensor in MoSSe Janus monolayer. MoSSe and WSSe Janus monolayers both belong to the $C_{3v}$ point group. Therefore, we expect the same qualitative trend, i.e. non-vanishing component of the out-of-plane SHG, also for WSSe Janus monolayer.


Finally, we discuss the second-order nonlinear dispersion at room and cryogenic temperatures in MoSSe and WSSe Janus monolayers, focusing on the in-plane component $\chi_{yyy}^{(2)}$ of the nonlinear susceptibility. For this purpose, we use a low-NA objective (NA = 0.3) to quench the non-planar components with negligible out-of-plane driving (see \hyperref[sec:SuppInfo]{SI} Figure \hyperref[sec:SuppInfo]{S3}e). Figures \ref{fig:Figure3}a and \ref{fig:Figure3}b show MoSSe Janus monolayer wavelength-dependent SHG spectra at room and cryogenic temperatures, respectively, obtained by sweeping the pumping energy, while maintaining constant pumping power. From the SHG intensities, we can derive the absolute values of the corresponding effective bulk-like second-order susceptibility dispersion $\chi_{yyy}^{(2)}$ \cite{woodward2016, autere2018}, presented in gray points and gray error band to account for excitation power fluctuations (see \hyperref[sec:SuppInfo]{SI} Section \hyperref[sec:SuppInfo]{S2}). For $\hbar\omega = \SI{0.885}{\electronvolt}$ excitation, we measure $\chi_{yyy}^{(2)}=\SI{51}{\pico\meter/\volt}$. Then, we straightforwardly obtain the other components $\chi_{xxz}^{(2)}=\chi_{zxx}^{(2)}=\SI{6}{\pico\meter/\volt}$, and $\chi_{zzz}^{(2)}=\SI{1.5}{\pico\meter/\volt}$, where  $\chi_{yyy}^{(2)}$ is highlighted in Figure \ref{fig:Figure3}a by the dotted line. 
Furthermore, when resonant with a real state, SHG is strongly enhanced \cite{wang2015} and it can thus be used as a background- and absorption-free spectroscopic tool to map the exciton states of Janus TMD monolayers. For the MoSSe Janus monolayer, we observe pronounced resonances corresponding to the $A$ and $B$ excitons, and note that the $A$ to $B$ exciton $\chi^{(2)}_{yyy}$ ratio evolves from the $B$ exciton dominating at room temperature to being the same as the $A$ exciton at $T = \SI{7}{\kelvin}$. This is in clear contrast to the WSSe Janus monolayer in Figures \ref{fig:Figure3}c and \ref{fig:Figure3}d at room and cryogenic temperatures, respectively. While $A$ and $B$ show similar values at $T = \SI{10}{\kelvin}$, $A$ dominates over $B$ at room temperature. In addition, the WSSe Janus monolayer dispersion shows additional resonances below $\SI{1.8}{\electronvolt}$ the origin of which remains unclear at present and requires further study. We tentatively ascribe them to charged excitons or defect bands. While we studied SHG dispersion in the $A$ and $B$ exciton range, another theoretically predicted resonance \cite{wei2019} has been observed in MoSSe monolayer close to \SI{2.75}{\electronvolt} \cite{bian2022}. 

In our experiments, the SH susceptibility of MoSSe and WSSe Janus monolayers at room temperature at exciton resonances is $\chi^{(2)}_{yyy}\approx\SI{150}{pm/V}$, and $\SI{200}{\pico\meter/\volt}$, respectively, which is among the largest reported values of second-order susceptibilities for any TMD monolayer \cite{lafeta2021}. In addition, we draw a qualitative and quantitative comparison between the measured dispersion in the MoSSe Janus monolayer and the corresponding calculations available in the literature \cite{wei2019, strasser2022, pike2022} (see \hyperref[sec:SuppInfo]{SI} Section \hyperref[sec:SuppInfo]{S5}). Reference \cite{pike2022} shows nearly dispersionless results in the exciton-dominated region, while references \cite{wei2019} and \cite{strasser2022} display two distinct resonances that align closely with the measured $A$ and $B$ excitons, providing a good qualitative agreement to the experimental data. Under quantitative comparison, experimental values consistently exhibit lower values compared to theoretical predictions. This is not surprising, as significant discrepancies are already reported between the calculated results (due to different calculation methods and corrections, excitonic effects, screening, temperature effects, etc.), while further reduction of the signal may be due to suboptimal sample quality.



\section*{Conclusion} 

In summary, we studied optical nonlinearities in MoSSe and WSSe Janus monolayers. We found an unusually low TH intensity in WSSe. In addition, we introduced a simple method to study their out-of-plane nonlinear susceptibility and measured values of $\chi_{xxz}^{(2)}=\chi_{zxx}^{(2)}=\SI{6}{\pico\meter/\volt}$ and $\chi_{zzz}^{(2)}=\SI{1.5}{\pico\meter/\volt}$ in MoSSe Janus monolayer. Finally, we provide the susceptibility dispersion on and off resonance across the exciton spectrum of both materials at room and cryogenic temperatures. Close to exciton resonances, we measured up to $\chi^{(2)}_{yyy}\approx\SI{200}{\pico\meter/\volt}$ for the in-plane SH nonlinear susceptibility, a value that is among the largest measured for atomically thin materials. Our work establishes a solid foundation to understand the nonlinear optical response of Janus TMD monolayers and their use in applications such as broadband frequency conversion and nonlinear photonic integrated circuits. Future steps should address two key challenges: (i) the comparatively low total efficiency; (ii) integration on-chip/fiber, which is one way to enhance the total nonlinear response and, thus, efficiency. However, in the latter, phase-matching conditions need to be considered. This has already been successfully demonstrated for conventional TMDs \cite{ngo2022}, while the direct growth of Janus TMD on-chip and photonic platforms is yet to be reported.

\section*{METHODS}
\label{sec:Methods}
\noindent\textbf{Samples.} Janus TMD monolayers were synthesized via a room temperature selective epitaxy atomic replacement method thoroughly explained in ref. \cite{trivedi2020}. MoSSe and WSSe Janus monolayers lay on Si/SiO$_{2}$ and Al$_{2}$O$_{3}$ substrates, respectively. The reference MoS$_2$ monolayer was encapsulated in thin hBN (HQ Graphene) and prepared on a Si/SiO$_2$ substrate with the same thickness as for MoSSe Janus monolayer. SHG maps from prototypical MoSSe and WSSe Janus monolayers were presented in \hyperref[sec:SuppInfo]{SI} Section \hyperref[sec:SuppInfo]{S6}. \\
\textbf{Optical Measurements.} 100-fs pulsed laser was used for SHG and THG measurements with a repetition rate of 80 MHz generated from an OPO (Levante IR fs, A.P.E.; and Inspire Femtosecond OPO, Radiantis) and pumped by a mode-locked Ti:sapphire laser, which was followed by a linear polarizer to ensure fixed excitation polarization.\\
For the polarization-resolved SHG measurements, a high-NA objective ($\mathrm{NA}=0.75$, Objective LD Plan-Neofluar 63x/0.75 Corr M27, Zeiss) was used  to reveal out-of-plane susceptibility tensor components. A longpass dichroic mirror was used to separate the generated SHG from the fundamental and direct it onto a single channel detector (silicon avalanche photodiode, APD440A, Thorlabs). Co-polarized excitation--detection configuration and polarization rotation were achieved by positioning a linear polarizer and a half-waveplate between the objective and the dichroic mirror. Detection noise, substrate influence, and acceptance angle range were taken into account.\\
For wavelength-dependent SHG measurements in Figure \ref{fig:Figure3}, and SHG and THG measurements in Figure \ref{fig:Figure1}, a low-NA achromatic objective ($\mathrm{NA}=0.3$, 15x Reflective microscope objective, Thorlabs) was used with a co-polarized excitation and detection polarizers. For cryogenic measurements a He-flow cryostat (Cryovac) was used. The collected light was analyzed in a spectrometer coupled to a charged-coupled device (Horiba).

\section*{Associated Content}
\subsection*{Supporting Information}
\label{sec:SuppInfo}
Selection rules for circularly polarized light and conservation of the angular momentum. Effective bulk-like susceptibility.  Polarization-resolved SHG intensity fitting function. Polarization-resolved SHG. Susceptibility dispersion comparison. SHG imaging of MoSSe and WSSe Janus monolayers.

\section*{Author Information}
\subsection*{Author Contributions}
M.M.P., G.S., J.J.F., and M.B. conceived and designed the experiment. Y.Q., Y.Sa., Y.Sh., and S.T. grew the Janus TMD monolayers, and A.B.M. fabricated the MoS$_2$ sample. M.M.P., V.V., P.H., and A.B.M. performed the optical measurements. M.M.P. analyzed the data and discussed them with all authors. M.M.P., G.S., and M.B. wrote the manuscript with input from all authors.
\subsection*{Notes}
The authors declare no competing financial interest.

\begin{acknowledgments}
S.T acknowledges primary support from DOE-SC0020653 (materials synthesis), Applied Materials Inc., NSF CMMI 1825594 (NMR and TEM studies), NSF DMR-1955889 (magnetic measurements), NSF CMMI-1933214, NSF 1904716, NSF 1935994, NSF ECCS 2052527, DMR 2111812, and CMMI 2129412. K.M. and J.J.F. acknowledge support from the European Union Horizon 2020 research and innovation programme under Grant Agreement No. 820423 (S2QUIP) and the Deutsche Forschungsgemeinschaft (DFG, German Research Foundation) under Germany’s Excellence Strategy – MCQST (EXC-2111), e-Conversion (EXC-2089), and via SPP 2299. M.M.P. acknowledges TUM International Graduate School of Science and Engineering (IGSSE). A.B.M. acknowledges funding from the International Max Planck Research School for Quantum Science and Technology (IMPRS-QST). M.B. and V.V. acknowledge the Alexander v. Humboldt foundation for financial support in the framework of their fellowship programs. K.M. acknowledges support from the Bayerische Akademie der Wissenschaften. G.S. acknowledges the German Research Foundation DFG (CRC 1375 NOA projects B5) and the IRTG 2675 META-ACTIVE (project A4).
\end{acknowledgments}


%

\end{document}



\title[\texttt{achemso} demonstration]
{SUPPORTING INFORMATION:\\Nonlinear Dispersion Relation and Out-of-Plane Second Harmonic Generation in\\ MoSSe and WSSe Janus Monolayers}


\author{Marko M. Petri\'{c}}
\email{Marko.Petric@wsi.tum.de}
\affiliation{Walter Schottky Institut and Department of Electrical and Computer Engineering, Technische Universit\"{a}t M\"{u}nchen, Am Coulombwall 4, 85748 Garching, Germany}
\affiliation{Munich Center for Quantum Science and Technology (MCQST), Schellingstrasse 4, 80799 Munich, Germany}

\author{Viviana Villafañe}
\affiliation{Walter Schottky Institut and Physik-Department, Technische Universit\"{a}t M\"{u}nchen, Am Coulombwall 4, 85748 Garching, Germany}
\affiliation{Munich Center for Quantum Science and Technology (MCQST), Schellingstrasse 4, 80799 Munich, Germany}

\author{Paul Herrmann}
\affiliation{Institute of Solid State Physics, Friedrich Schiller University Jena, Max-Wien-Platz 1, 07743 Jena, Germany}

\author{Amine Ben Mhenni}
\affiliation{Walter Schottky Institut and Physik-Department, Technische Universit\"{a}t M\"{u}nchen, Am Coulombwall 4, 85748 Garching, Germany}
\affiliation{Munich Center for Quantum Science and Technology (MCQST), Schellingstrasse 4, 80799 Munich, Germany}

\author{Ying Qin}
\affiliation{Materials Science and Engineering, School for Engineering of Matter, Transport and Energy, Arizona State University, Tempe, Arizona 85287, USA}

\author{Yasir Sayyad}
\affiliation{Materials Science and Engineering, School for Engineering of Matter, Transport and Energy, Arizona State University, Tempe, Arizona 85287, USA}

\author{Yuxia Shen}
\affiliation{Materials Science and Engineering, School for Engineering of Matter, Transport and Energy, Arizona State University, Tempe, Arizona 85287, USA}

\author{Sefaattin Tongay}
\email{sefaattin.tongay@asu.edu}
\affiliation{Materials Science and Engineering, School for Engineering of Matter, Transport and Energy, Arizona State University, Tempe, Arizona 85287, USA}

\author{Kai M\"{u}ller}
\affiliation{Walter Schottky Institut and Department of Electrical and Computer Engineering, Technische Universit\"{a}t M\"{u}nchen, Am Coulombwall 4, 85748 Garching, Germany}
\affiliation{Munich Center for Quantum Science and Technology (MCQST), Schellingstrasse 4, 80799 Munich, Germany}

\author{Giancarlo Soavi}
\email{giancarlo.soavi@uni-jena.de}
\affiliation{Institute of Solid State Physics, Friedrich Schiller University Jena, Max-Wien-Platz 1, 07743 Jena, Germany}
\affiliation{Abbe Center of Photonics, Friedrich Schiller University Jena, Albert-Einstein-Straße 6, 07745 Jena, Germany\\}

\author{Jonathan J. Finley}
\affiliation{Walter Schottky Institut and Physik-Department, Technische Universit\"{a}t M\"{u}nchen, Am Coulombwall 4, 85748 Garching, Germany}
\affiliation{Munich Center for Quantum Science and Technology (MCQST), Schellingstrasse 4, 80799 Munich, Germany}

\author{Matteo Barbone}
\email{Matteo.Barbone@wsi.tum.de}
\affiliation{Walter Schottky Institut and Department of Electrical and Computer Engineering, Technische Universit\"{a}t M\"{u}nchen, Am Coulombwall 4, 85748 Garching, Germany}
\affiliation{Munich Center for Quantum Science and Technology (MCQST), Schellingstrasse 4, 80799 Munich, Germany}

\maketitle
\onecolumngrid

\section{S1. Selection rules for circularly polarized light and conservation of the angular momentum}

When exciting a crystal with a circularly polarized fundamental beam at normal incidence, conservation of out-of-plane angular momentum allows for transitions that obey: 
\begin{equation}
\Delta L_{\mathrm{photon}} = \Delta L_{\mathrm{valley}} + \Delta L_{\mathrm{exciton}} + \Delta L_{\mathrm{lattice}}.
\label{eq:eq1}
\end{equation} 
Here, $\Delta L_{\mathrm{photon}} = \Delta m\hbar$, $\Delta L_{\mathrm{valley}} = \Delta\tau\hbar$, $\Delta L_{\mathrm{exciton}} = \Delta l\hbar$, and $\Delta L_{\mathrm{lattice}}=\pm3n\hbar$ (for crystals with a three-fold rotational symmetry) stand for changes in angular momenta of photons, valley, excitons, and crystal lattice, respectively \cite{S-bloembergen1980, S-xiao2015}. In the case of the $1s$-exciton resonant SHG, exciton angular momentum does not flip, while the valley angular momentum has zero net change upon transition of the system from the ground state to the excited state and back; thus, Eq. (\ref{eq:eq1}) can be simplified to $\Delta m\hbar = \pm3\hbar$ ($n=1$). The left side represents the net change of the spin of the photons, while the right side represents the angular momentum transferred to the crystal, which comes in integer number $n$ of $\pm3\hbar$ as a direct result of the three-fold crystal symmetry \cite{S-bloembergen1980, S-xiao2015}. Therefore, SHG emission always results in the opposite helicity of the incident light since $L_{\mathrm{out}\sigma^{\pm}} - 2L_{\mathrm{in}\sigma^{\mp}} = \pm3\hbar$, where $L_{\mathrm{in}}$ and $L_{\mathrm{out}}$ stand for angular momenta of incoming and outgoing photons, and $L_{\sigma^{\pm}} = \pm\hbar$. 

\begin{figure}[h]
\centering
\includegraphics{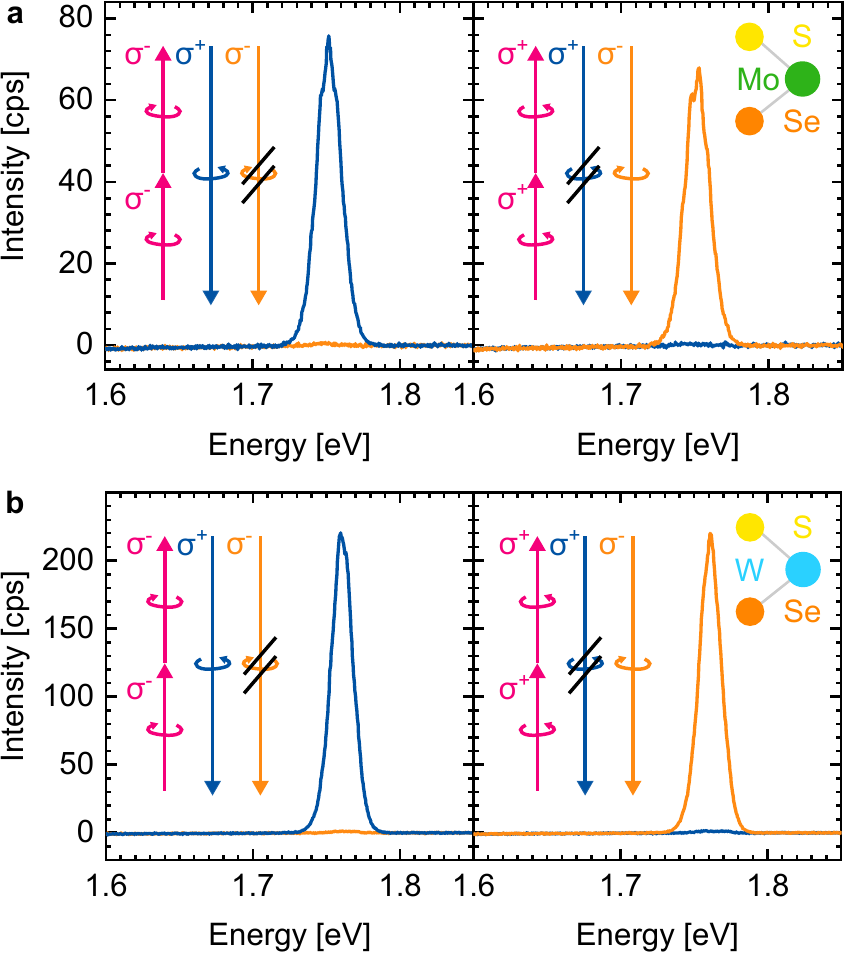}
\caption{\textbf{SHG polarization selection rules.} \textbf{(a,b)} SHG spectra with $\sigma^+$ (dark blue) and $\sigma^-$ (orange) circular-polarization filtering in the detection upon $\sigma^-$ (left panel) and $\sigma^+$ (right panel) two-photon excitation (magenta arrows) for \textbf{(a)} MoSSe and \textbf{(b)} WSSe Janus monolayers.}
\label{fig:FigurePol}
\end{figure}

This behavior holds for MoSSe and WSSe monolayers at room temperature as presented in Figures \ref{fig:FigurePol}a and \ref{fig:FigurePol}b, respectively. Here, left (right) panels show that pumping a left (right) circularly polarized light results in a strong right (left) circularly polarized SH signal and a highly suppressed left (right) circularly polarized SH signal. We tuned the two-photon energy above the real exciton transition for MoSSe and resonant to it for WSSe Janus monolayer. Measured SHG intensities $I_{\sigma^{\pm}}$ result in a high degree of circular polarization $|(I_{\sigma^{\pm}} - I_{\sigma^{\mp}})/(I_{\sigma^{\pm}} + I_{\sigma^{\mp}})| = 96-99\%$, which is due to a parametric nature of the SHG, which is robust to the valley scattering \cite{S-xiao2015, S-seyler2015, S-zhang2020}. Furthermore, this behavior should hold irrespective of the excitation energy, as the selection rule $\Delta m\hbar = \pm3\hbar$ is ultimately independent of the valley degree of freedom and stems solely from the three-fold crystal symmetry, a property retained in all TMD monolayers.

\section{S2. Effective bulk-like susceptibility}

As demonstrated in Figures 1e and 1f (Figure 1i and Figure 1j) in the main text, SHG (THG) intensity $P_{\mathrm{SHG}}(2\omega)$ ($P_{\mathrm{THG}}(3\omega)$) has quadratic (cubic) dependence with respect to the pump power $P_1(\omega)$ \cite{S-woodward2016}:

\begin{equation}
    P_{\mathrm{SHG}}(2\omega) = \frac{16\sqrt{2}S\big|\chi^{(2)}\big|^2\omega^2}{c^3\epsilon_0^2f\pi r^2t_{\mathrm{fwhm}}(1+n_2)^6}P_1^2(\omega)
    \label{eq:P_SHG}
\end{equation}

\begin{equation}
    P_{\mathrm{THG}}(3\omega) = \frac{64\sqrt{3}S^2\big|\chi^{(3)}\big|^2\omega^2}{c^4\epsilon_0^2(ft_{\mathrm{fwhm}}\pi r^2)^2(1+n_2)^8}P_1^3(\omega)
    \label{eq:P_THG}
\end{equation}

Here, $\omega$ is the pumping angular frequency, $S=0.94$ is a Gaussian pulse shape factor, $f=\SI{80}{MHz}$ is the pump laser repetition rate, $t_{\mathrm{fwhm}} =\SI{100}{\femto\second}$ is the pulse length, $n_2$ is the substrate refractive index, $c$ is the speed of light in vacuum, $\epsilon_0$ is the permittivity of free-space, and $\pi r^2$ is the laser spot area. Second-order susceptibility $\chi^{(2)}$ and third-order susceptibility $\chi^{(3)}$ can be thus derived from Eqs. \ref{eq:P_SHG} and \ref{eq:P_THG}, where SHG and THG power can be extracted from the area of measured spectral peaks. Finally, in the main text, we provide values of effective bulk-like susceptibilities (Figure 3 in the main text for $\chi_{yyy}^{(2)}$ and Figures 1g and 1h for $\chi_{\mathrm{eff}}^{(3)}$) common in the literature, which are related to the susceptibilities in Eqs. \ref{eq:P_SHG} and \ref{eq:P_THG} as \cite{S-woodward2016, S-autere2018}:

\begin{equation}
    \big|\chi_{\mathrm{eff}}^{(2,3)}\big| = \big|\chi^{(2,3)}\big|/h
\end{equation}

where $h$ is a monolayer thickness ($h_{\mathrm{MoSSe}}=\SI{0.65}{\nano\meter}$). To finally obtain $\chi^{(2)}$ values presented in the main text, we also take into account the efficiency of the setup.

\section{S3. Polarization-resolved SHG intensity fitting function}

To arrive at different SHG intensities $I_{x,y,z}$ for the specific measurement configuration, we start from Eq. (2) in the main text and transform the incident electric field to polar coordinates. The resulting expressions are equivalent to the set of equations (9) in ref. \cite{S-pike2022} for polarization angle $\alpha = 0$, corresponding to the co-polarized configuration we use in our measurements:

\begin{equation}
    I_x\propto\big|\chi_{xxz}^{(2)}\sin 2\theta + \chi_{yyy}^{(2)}\cos^2 \theta\textcolor{cyan}{\sin 3\phi}\big|^2
    \label{eq:Ix}
\end{equation}
\begin{equation}
    I_y\propto\big|\chi_{yyy}^{(2)}\cos^2 \theta\cos 3\phi\big|^2
    \label{eq:Iy}
\end{equation}
\begin{equation}
    I_z\propto\big|\chi_{zxx}^{(2)}\cos^2\theta + \chi_{zzz}^{(2)}\sin^2\theta\big|^2
    \label{eq:Iz}
\end{equation}

which are equivalent to Eqs. (3)--(5) in the main text. However, these intensities do not take substrate effects into account. Furthermore, they only predict values for a specific excitation angle $\theta$ and not the full range within the objective acceptance angle. In the following lines, we incorporate these two points to Eqs. \ref{eq:Ix}, \ref{eq:Iy}, and \ref{eq:Iz} to obtain the measured intensity $I_p$ probed in our experiment.

\begin{figure}[h]
\centering
\includegraphics{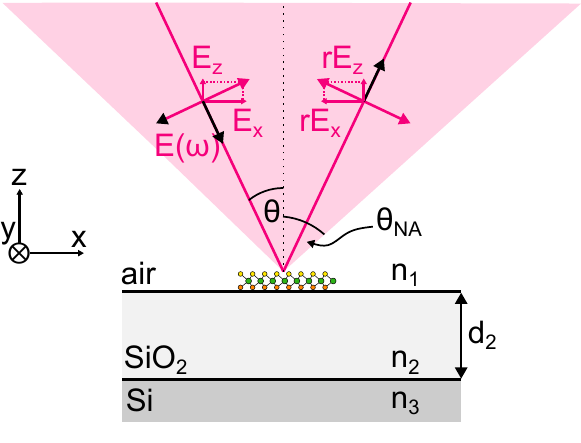}
\caption{\textbf{Incoming light field $\mathbf{E(\omega)}$ from the objective focused onto MoSSe Janus monolayer.} Magenta denotes light coming by an angle $\theta$ from the optical axis (dotted line), limited to $\theta_{\mathrm{NA}}$ by the objective numerical aperture. Refractive indices $n_{1,2,3}$ stand for air, SiO$_2$ (of thickness $d_2 = \SI{285}{\nano\meter}$) and Si (quasi-infinite thickness), respectively. Total reflection coefficient $r$ can be calculated from Eq. S9. 
}
\label{fig:Fresnel}
\end{figure}

The electric field impinging on the MoSSe Janus monolayer interferes with the reflected light from the substrate; thus, causing the following input electric field components, sketched in Figure \ref{fig:Fresnel} \cite{S-lu2017}:

\begin{align}
    &E_x\rightarrow\textcolor{magenta}{E_x(1-r)}\\ \label{eq:inputX}
    &E_z\rightarrow\textcolor{magenta}{E_z(1+r)}\\ \label{eq:inputZ}
    &\textcolor{magenta}{r} = \frac{r_{12} + \exp(i2\delta)r_{23}}{1 + \exp(i2\delta)r_{12}r_{23}} 
\end{align}

Here, $\textcolor{magenta}{r}$ is the total reflection coefficient for three different media in Figure \ref{fig:Fresnel}, where we neglect the effect from the monolayer due to its small thickness; ${r_{ij}}$ are reflection coefficients between interfaces $i$ and $j$ ($i,j = 1,2,3$), $\delta$ is the accumulated phase in the medium $2$ (SiO$_2$) of thickness $d_2 = \SI{285}{\nano\meter}$ for the light of wavelength $\lambda = \SI{1400}{\nano\meter}$. Coefficients $r_{ij}$ and phase $\delta$ can be explicitly written as:

\begin{equation}
    \delta = \frac{2\pi d_2}{\lambda}n_2\cos\theta_2
\end{equation}

\begin{equation}
    r_{ij} = \frac{n_j\cos\theta_i - n_i\cos\theta_j}{n_j\cos\theta_i + n_i\cos\theta_j}
    \label{eq:Fresnel}
\end{equation}

Eq. \ref{eq:Fresnel} is one of the well-known Fresnel equations, where $n_{i,j}$ and $\theta_{i,j}$ are refractive indices and light incidence angles with respect to the $z$-axis, respectively. Different refraction angles are connected \textit{via} Snell-Descartes law:

\begin{equation}
    n_i\cos\theta_i = n_j\cos\theta_j
\end{equation}

By performing matrix multiplication in Eq. (2) in the main text and incorporating new values for the input fields in Eqs. \ref{eq:inputX} and \ref{eq:inputZ}, we arrive at more accurate expressions for $I_{x,y,z}$:

\begin{equation}
    I_x\propto\big|\chi_{xxz}^{(2)}\textcolor{magenta}{(1-r)(1+r)}\sin 2\theta + \chi_{yyy}^{(2)}\textcolor{magenta}{(1-r)^2}\cos^2 \theta\textcolor{cyan}{\sin 3\phi}\big|^2
    \label{eq:rIx}
\end{equation}

\begin{equation}
    I_y\propto\big|\chi_{yyy}^{(2)}\textcolor{magenta}{(1-r)^2}\cos^2 \theta\cos 3\phi\big|^2
    \label{eq:rIy}
\end{equation}

\begin{equation}
    I_z\propto\big|\chi_{zxx}^{(2)}\textcolor{magenta}{(1-r)^2}\cos^2\theta + \chi_{zzz}^{(2)}\textcolor{magenta}{(1+r)^2}\sin^2\theta\big|^2
    \label{eq:rIz}
\end{equation}

We highlight \textcolor{magenta}{reflection-induced terms in magenta} and \textcolor{cyan}{$\phi$-dependent terms in cyan} throughout the rest of the derivation for clarity. The objective collects both components $I_x$ and $I_z$ in the plane of incidence; thus, the detected light has the intensity $I_p$:

\begin{equation}
    I_p = I_x\cos^2\theta + I_z\sin^2\theta
    \label{eq:Ip_initial}
\end{equation}

Inserting \ref{eq:rIx} and \ref{eq:rIz} into \ref{eq:Ip_initial}, we obtain:

\begin{equation}
    I_p\propto\big|\chi_{xxz}^{(2)}\cdot2\sin\theta\cos^2\theta\textcolor{magenta}{(1-r)(1+r)} + \chi_{yyy}^{(2)}\cos^3\theta\textcolor{magenta}{(1-r)^2}\textcolor{cyan}{\sin3\phi}\big|^2 + \big|\chi_{zxx}^{(2)}\sin\theta\cos^2\theta\textcolor{magenta}{(1-r)^2} + \chi_{zzz}^{(2)}\sin^3\theta\textcolor{magenta}{(1+r)^2}\big|^2
\end{equation}

This expression holds for a specific angle $\theta$. However, the objective accepts light of all angles between $0$ and $\arcsin\mathrm{(NA)}$, where $\mathrm{NA} = 0.75$ is the numerical aperture. Therefore, we integrate $I_p$ over this angle range to obtain the measured $I_p$:

\begin{equation}
    I_p\propto\int_{0}^{\arcsin(\mathrm{NA})}(\big|\underbrace{\chi_{zxx}^{(2)}\textcolor{ForestGreen}{q_1} + \chi_{zzz}^{(2)}\textcolor{ForestGreen}{q_2}}_{\rightarrow\mathbb{A}}\big|^2 + \big|\underbrace{\chi_{xxz}^{(2)}\textcolor{ForestGreen}{q_3} + \chi_{yyy}^{(2)}\textcolor{ForestGreen}{q_4}\textcolor{cyan}{\sin3\phi}\big|^2}_{\rightarrow\mathbb{B}})\mathrm{d}\theta
    \label{eq:Ip_int}
\end{equation}

Here, we introduce $\theta$-dependent expressions $\textcolor{ForestGreen}{q_{1,2,3,4}}$ for readability, highlighted in \textcolor{ForestGreen}{green}:

\begin{align}
    & \textcolor{ForestGreen}{q_1} = \sin\theta\cos^2\theta\textcolor{magenta}{(1-r)^2}\\
    & \textcolor{ForestGreen}{q_2} = \sin^3\theta\textcolor{magenta}{(1+r)^2}\\
    & \textcolor{ForestGreen}{q_3} = 2\sin\theta\cos^2\theta\textcolor{magenta}{(1-r)(1+r)}\\
    & \textcolor{ForestGreen}{q_4} = \cos^3\theta\textcolor{magenta}{(1-r)^2}
\end{align}

We further develop integrals $\mathbb{A}$ and $\mathbb{B}$, where we implicitly use integration boundaries as defined in \ref{eq:Ip_int}:

\begin{align}
    \nonumber
    \mathbb{A} &= \int\big|\chi_{zxx}^{(2)}\textcolor{ForestGreen}{q_1} + \chi_{zzz}^{(2)}\textcolor{ForestGreen}{q_2}\big|^2\mathrm{d}\theta = \int(\chi_{zxx}^{(2)}\textcolor{ForestGreen}{q_1} + \chi_{zzz}^{(2)}\textcolor{ForestGreen}{q_2})(\chi_{zxx}^{(2)}\textcolor{ForestGreen}{q_1^*} + \chi_{zzz}^{(2)}\textcolor{ForestGreen}{q_2^*})\mathrm{d}\theta\\ \nonumber
    &= \int(\big|\chi_{zxx}^{(2)}\big|^2\textcolor{ForestGreen}{q_1q_1^*} + \big|\chi_{zzz}^{(2)}\big|^2\textcolor{ForestGreen}{q_2q_2^*} + \chi_{zxx}^{(2)}\chi_{zzz}^{(2)}(\textcolor{ForestGreen}{q_1^*q_2 + q_1q_2^*}))\mathrm{d}\theta\\ 
    &= \textcolor{RedOrange}{\mathcal{I}_{11}}\big|\chi_{zxx}^{(2)}\big|^2 + \textcolor{RedOrange}{\mathcal{I}_{22}}\big|\chi_{zzz}^{(2)}\big|^2 + \textcolor{RedOrange}{\mathcal{I}_{12}}\chi_{zxx}^{(2)}\chi_{zzz}^{(2)}
\end{align}

\begin{align}
    \nonumber
    \mathbb{B} &= \int\big|\chi_{xxz}^{(2)}\textcolor{ForestGreen}{q_3} + \chi_{yyy}^{(2)}\textcolor{ForestGreen}{q_4}\textcolor{cyan}{\sin3\phi}\big|^2 \mathrm{d}\theta = \int(\chi_{xxz}^{(2)}\textcolor{ForestGreen}{q_3} + \chi_{yyy}^{(2)}\textcolor{ForestGreen}{q_4}\textcolor{cyan}{\sin3\phi})(\chi_{xxz}^{(2)}\textcolor{ForestGreen}{q_3^*} + \chi_{yyy}^{(2)}\textcolor{ForestGreen}{q_4^*}\textcolor{cyan}{\sin3\phi})\mathrm{\theta} \\ \nonumber
    &= \int(\big|\chi_{xxz}^{(2)}\big|^2\textcolor{ForestGreen}{q_3q_3^*} + \big|\chi_{yyy}^{(2)}\big|^2\textcolor{ForestGreen}{q_4q_4^*}\textcolor{cyan}{\sin^23\phi} + \chi_{xxz}^{(2)}\chi_{yyy}^{(2)}(\textcolor{ForestGreen}{q_3q_4^* + q_3^*q_4})\textcolor{cyan}{\sin3\phi} )\mathrm{d}\theta\\
    &= \textcolor{RedOrange}{\mathcal{I}_{33}}\big|\chi_{xxz}^{(2)}\big|^2 + \textcolor{RedOrange}{\mathcal{I}_{44}}\big|\chi_{yyy}^{(2)}\big|^2\textcolor{cyan}{\sin^23\phi} + \textcolor{RedOrange}{\mathcal{I}_{34}}\chi_{xxz}^{(2)}\chi_{yyy}^{(2)}\textcolor{cyan}{\sin3\phi}
\end{align}

Here, we introduce the following integrals in \textcolor{RedOrange}{orange} which we later evaluate numerically:

\begin{align}
    & \textcolor{RedOrange}{\mathcal{I}_{11}} = \int \textcolor{ForestGreen}{q_1q_1^*}\mathrm{d}\theta = \int\sin^2\theta\cos^4\theta\textcolor{magenta}{(1-r)^2(1-r^*)^2}\mathrm{d}\theta\\
    & \textcolor{RedOrange}{\mathcal{I}_{22}} = \int \textcolor{ForestGreen}{q_2q_2^*}\mathrm{d}\theta = \int\sin^6\theta\textcolor{magenta}{(1+r)^2(1+r^*)^2}\mathrm{d}\theta\\
    & \textcolor{RedOrange}{\mathcal{I}_{12}} = \int \textcolor{ForestGreen}{(q_1q_2^*+q_1^*q_2)}\mathrm{d}\theta = \int\sin^4\theta\cos^2\theta\textcolor{magenta}{((1-r)^2(1+r^*)^2 + (1-r^*)^2(1+r)^2)}\mathrm{d}\theta\\
    & \textcolor{RedOrange}{\mathcal{I}_{33}} = \int \textcolor{ForestGreen}{q_3q_3^*}\mathrm{d}\theta = \int4\sin^2\theta\cos^4\theta\textcolor{magenta}{(1-r^2)(1-r^{*2})}\mathrm{d}\theta\\
    & \textcolor{RedOrange}{\mathcal{I}_{44}} = \int \textcolor{ForestGreen}{q_4q_4^*}\mathrm{d}\theta = \int\cos^6\theta\textcolor{magenta}{(1-r)^2(1-r^*)^2}\mathrm{d}\theta\\
    & \textcolor{RedOrange}{\mathcal{I}_{34}} = \int \textcolor{ForestGreen}{(q_3q_4^*+q_3^*q_4)}\mathrm{d}\theta = \int2\sin\theta\cos^5\theta\textcolor{magenta}{((1-r^2)(1-r^*)^2 + (1-r^{*2})(1-r)^2)}\mathrm{d}\theta
\end{align}

Using the above defined integrals, we can rewrite Eq. \ref{eq:Ip_int} as:

\begin{equation}
I_p\propto\mathbb{A} + \mathbb{B} = \textcolor{RedOrange}{\mathcal{I}_{11}}\big|\chi_{zxx}^{(2)}\big|^2 + \textcolor{RedOrange}{\mathcal{I}_{22}}\big|\chi_{zzz}^{(2)}\big|^2 + \textcolor{RedOrange}{\mathcal{I}_{12}}\chi_{zxx}^{(2)}\chi_{zzz}^{(2)} + 
\textcolor{RedOrange}{\mathcal{I}_{33}}\big|\chi_{xxz}^{(2)}\big|^2 + \textcolor{RedOrange}{\mathcal{I}_{44}}\big|\chi_{yyy}^{(2)}\big|^2\textcolor{cyan}{\sin^23\phi} + \textcolor{RedOrange}{\mathcal{I}_{34}}\chi_{xxz}^{(2)}\chi_{yyy}^{(2)}\textcolor{cyan}{\sin3\phi}
\label{eq:AlmostFinal}
\end{equation}

As discussed in the main text, to give a quantitative estimate of the out-of-plane components compared to the in-plane components, we reduce the number of unknown parameters in Eq. \ref{eq:AlmostFinal}, and thus, simplify the $\boldsymbol\chi^{(2)}$ tensor by applying the the Kleinman's symmetry to obtain $\chi_{xxz}^{(2)} = \chi_{zxx}^{(2)}$ \cite{S-boyd2008a, S-lu2017}. While in ref. \cite{S-lu2017} $\chi_{zzz}^{(2)}$ was neglected, we use a more reasonable estimate consulting theory \cite{S-pike2022} to further reduce the parameter space by expressing $\chi_{zzz}^{(2)}$ in terms of $\chi_{xxz}^{(2)}$, where $\chi_{xxz}^{(2)}/\chi_{zzz}^{(2)}\approx4$. Finally, we come to the final expression containing only parameters $\mathbb{P}_1$ and $\mathbb{P}_2$:

\begin{equation}
    \boxed{
    I_p\propto
    (\textcolor{RedOrange}{\mathcal{I}_{11}} + \frac{1}{16}\textcolor{RedOrange}{\mathcal{I}_{22}}+\frac{1}{4}\textcolor{RedOrange}{\mathcal{I}_{12}}+\textcolor{RedOrange}{\mathcal{I}_{33}}){\underbrace{\big|\chi_{xxz}^{(2)}\big|}_{\mathbb{P}_1}}^2+
    \textcolor{RedOrange}{\mathcal{I}_{34}}\underbrace{\chi_{xxz}^{(2)}}_{\mathbb{P}_1}\underbrace{\chi_{yyy}^{(2)}}_{\mathbb{P}_2}\textcolor{cyan}{\sin3\phi}+
    \textcolor{RedOrange}{\mathcal{I}_{44}}{\underbrace{\big|\chi_{yyy}^{(2)}\big|}_{\mathbb{P}_2}}^2\textcolor{cyan}{\sin^23\phi}
    }
    \label{eq:FINAL}
\end{equation}

We calculate the integrals numerically to obtain the following values:

\begin{equation}
    \textcolor{RedOrange}{\mathcal{I}_{11}} = 0.1201,\;\textcolor{RedOrange}{\mathcal{I}_{22}} = 0.1161,\;\textcolor{RedOrange}{\mathcal{I}_{12}} = -0.1016,\;\textcolor{RedOrange}{\mathcal{I}_{33}} = 0.5924,\;\textcolor{RedOrange}{\mathcal{I}_{44}} = 1.1631,\;\textcolor{RedOrange}{\mathcal{I}_{34}} = 0.5002.
\end{equation}

Finally, by fitting the data in Figure 2 in the main text using \ref{eq:FINAL} and relative weighting, we extract the parameters $\mathbb{P}_1$ and $\mathbb{P}_2$, and thus, obtain the ratio $\boxed{\mathbb{P}_2/\mathbb{P}_1 = \chi_{yyy}^{(2)}/\chi_{xxz}^{(2)} = 8.4}$.

\section{S4. Polarization-resolved SHG}

\begin{figure}[h]
\centering
\includegraphics{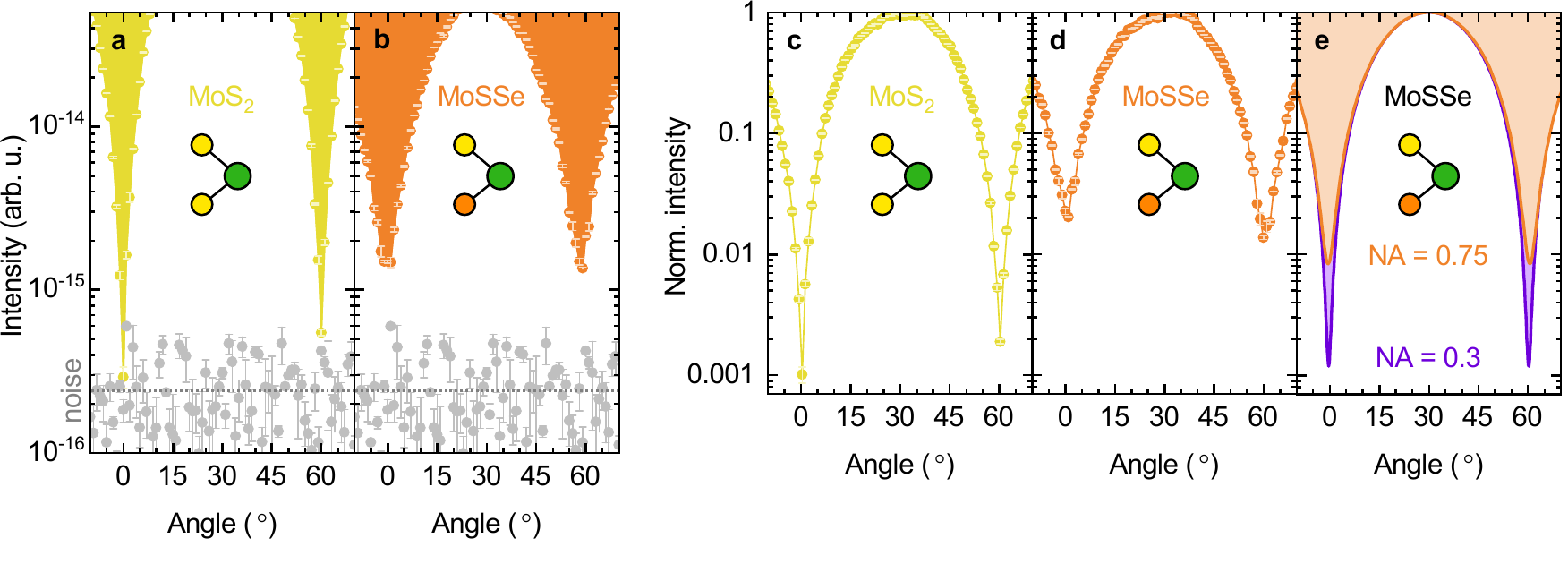}
\caption{\textbf{Polarization-resolved SHG intensity of $\mathbf{MoSSe}$ Janus monolayer and $\mathbf{MoS}_2$ monolayer} \textbf{(a,b)} Absolute intensity of polarization-resolved SHG of \textbf{(a)} MoS$_2$ (yellow) and \textbf{(b)} MoSSe Janus (orange) monolayer compared to the APD dark counts. \textbf{(c,d)} Logarithmic plots of the normalized polarization-resolved SHG intensity of \textbf{(c)} MoS$_2$ (yellow) and \textbf{(d)} MoSSe Janus (orange) monolayer. Error bars represent standard deviation obtained from multiple readouts. \textbf{(e)} Comparison between the measurements with two different objectives. NA = 0.75 (orange) shows a pronounced out-of-plane response that can be easily distinguished from the underlying noise, while the signal reaches the noise floor with NA = 0.3 (purple) comparable to an MoS$_2$ monolayer in \textbf{(c)}.}
\label{fig:Noise}
\end{figure}

To obtain polarization-resolved SHG intensity data, we first substract the instrumentation noise (dark counts), shown in Figure \ref{fig:Noise}a and \ref{fig:Noise}b in gray. To eliminate the influence of potential setup-induced polarization distortions, we perform a control experiment on a well-studied MoS$_2$ monolayer and subsequently compare it to MoSSe Janus monolayer response, while keeping the same laser wavelength and power. For both materials, excitation is far from the exciton resonances to avoid unwanted TP-PL signal. Figure \ref{fig:Noise}b shows that the minimum detected SHG intensity of MoSSe Janus monolayer is well above the noise level, in constrast to the signal from MoS$_2$ monolayer which shows minima within the noise band as presented in \ref{fig:Noise}a. Figure \ref{fig:Noise}c shows noise-corrected normalized SHG intensity of MoS$_2$, reaching minima one order of magnitude smaller than MoSSe Janus monolayer in Figure \ref{fig:Noise}d.

To determine how the minimum-to-maximum SHG intensity ratio changes with the acceptance angle, we also performed polarization-dependent SHG measurements using NA = 0.3 objective. Here, we fit the data to the model analogous to Eq. \ref{eq:FINAL}, only integrated over the angle $\theta$ from $0$ to $\arcsin(\mathrm{NA} = 0.3)$, and subsequently compare it to the NA = 0.75 case (Eq. \ref{eq:FINAL} integrated from $0$ to $\arcsin(\mathrm{NA} = 0.75)$), while keeping the same $\chi^{(2)}$ ratios. The results are presented in Figure \ref{fig:Noise}e. The minimum-to-maximum intensity ratios are significantly decreased in the case of NA = 0.3 ($R_{0.3} = I_{\mathrm{min}}/I_{\mathrm{max}} \sim 1/750$)  compared to NA = 0.75 ($R_{0.75} = I_{\mathrm{min}}/I_{\mathrm{max}} \sim 1/120$), resulting in $R_{0.3}/R_{0.75} \sim 1/6.2$. The signal obtained from NA = 0.3 is already close to the noise level and cannot be distinguished from the MoS$_2$ monolayer (compare with Figure \ref{fig:Noise}c), introducing large errors in the extracted parameters, and thus making it difficult to resolve the effect. Furthermore, this is in accordance with the theoretical estimate of the minimum-to-maximum ratios between different excitation/acceptance angles where $R_{0.3}/R_{0.75} \sim 1/6$, as derived from Eq. \ref{eq:FINAL}.

\section{S5. Susceptibility dispersion comparison}

\begin{figure}[h]
\centering
\includegraphics{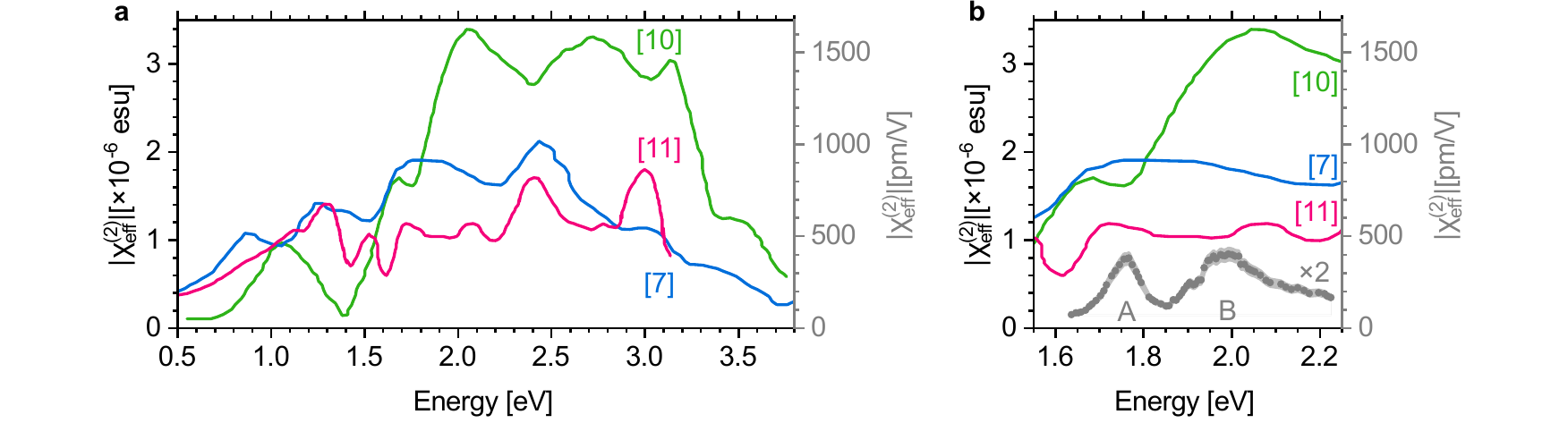}
\caption{\textbf{Effective in-plane susceptibility dispersion of a MoSSe monolayer.} \textbf{(a,b)} Effective in-plane susceptibility dispersions of a MoSSe monolayer from three different theoretical works: ref. \cite{S-pike2022} (blue), ref. \cite{S-wei2019} (green), and ref. \cite{S-strasser2022} (magenta) (references [55, 53, 54] in the main text, respectively). The left \textit{y}-axis is in the units of $10^{-6}$ esu, and the right \textit{y}-axis in the units of pm/V. \textbf{(b)} Susceptibility dispersion comparison in the energy region of the experimental data, represented in gray, with pronounced A and B excitons (scaled with a factor of two for clarity).
}
\label{fig:Compare}
\end{figure}

Figure \ref{fig:Compare}a presents the calculated in-plane susceptibility dispersions of a MoSSe monolayer from three different theoretical works \cite{S-wei2019, S-strasser2022, S-pike2022} (references [55, 53, 54] in the main text, respectively). To the best of our knowledge, these are the only such works present in the literature so far. For clarity, we represent all effective $\chi_{yyy}^{(2)}$ values in the units of $10^{-6}$ esu (left \textit{y}-axis) and in the units of pm/V (right \textit{y}-axis). The results differ qualitatively and quantitatively in most of the dispersion region. 

There are several reasons for this. First, different calculation methods and corrections were used. Second, models assume parameters and assumptions that are empirically unknown due to lack of experimental results. Third, quantitative assessment of excitonic effects remains difficult to calculate, where only ref. \cite{S-wei2019} uses GW-BSE methods. Bare bare DFT produces a significant error when calculating the absolute values of excitonic energies. Moreover, substrate screening effects and temperature dependence are important realistic parameters that are not taken into account. Therefore, the quantitative (to a larger extent), but also qualitative discrepancies are to be expected.

Figure \ref{fig:Compare}b depicts the comparison of different theoretical results to the experiment at 7 K for a MoSSe monolayer confined to the measured spectral window. 

\textit{Qualitative comparison}: the results from ref. \cite{S-pike2022} appear to be almost dispersionless in this region where excitons should dominate, while the results  from ref. \cite{S-wei2019} and ref. \cite{S-strasser2022} exhibit two distinct resonances spectrally close to measured A and B excitons, rendering good qualitative correspondence to the measured data. The first (second) resonance in ref. \cite{S-wei2019} and ref. \cite{S-strasser2022} agrees with the A (B) exciton for the measured data at 7 K up to 0.1 eV. In addition, it is worth mentioning that the resonance close to 2.75 eV in Figure \ref{fig:Compare}a ref. \cite{S-wei2019} has been observed in \cite{S-bian2022} (ref. [44] in the main text).

\textit{Quantitative comparison}: Experimental results consistently show lower values compared to the calculations. At exciton resonances, we estimate experimental value $\sim$200 pm/V, while theoretical values are $\sim$500 pm/V in ref. \cite{S-strasser2022}, $\sim$800 pm/V in ref. \cite{S-pike2022}, and $\sim$800 pm/V (A exciton) and $\sim$1600 pm/V (B exciton) in ref. \cite{S-wei2019}. Therefore, we record $\sim$2.5-4 times for A exciton, and $\sim$2.5-8 times for B exciton lower values than predicted by theory. This is also not surprising, as significant discrepancies exist already within the calculated results, and further reduction of the signal may be due to suboptimal sample quality.

\section{S6. SHG imaging of $\mathbf{MoSSe}$ and $\mathbf{WSSe}$ Janus monolayers}

\begin{figure}[h]
\centering
\includegraphics{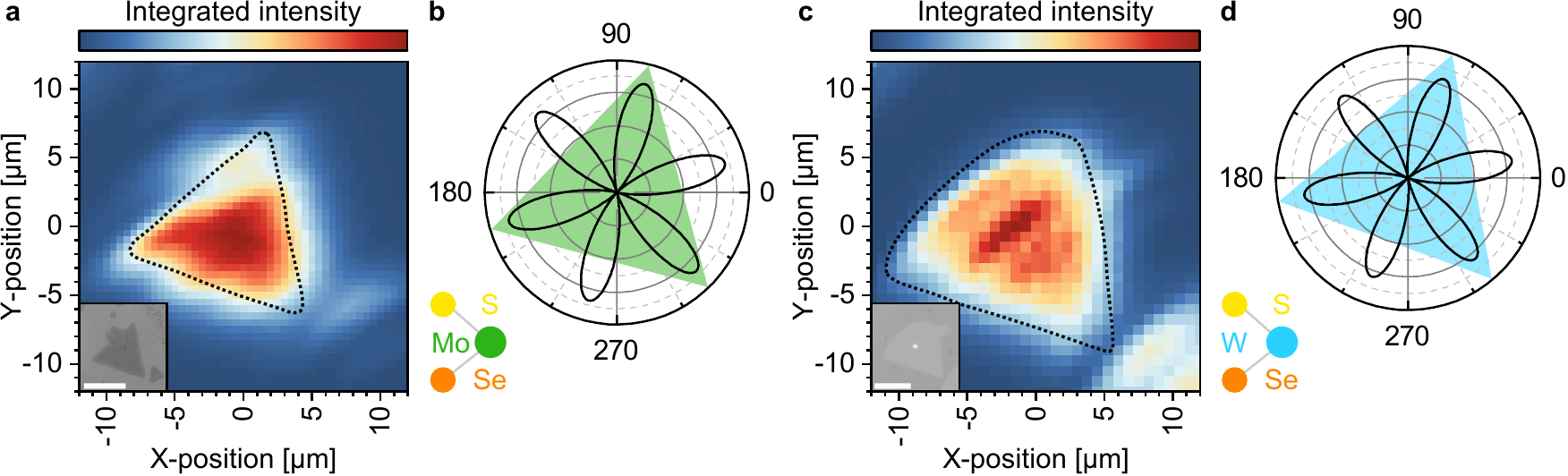}
\caption{\textbf{SHG imaging of MoSSe and WSSe Janus monolayer flakes.} \textbf{(a,c)} Spatial map of integrated SHG intensity from \textbf{(a)} MoSSe and \textbf{(c)} WSSe Janus monolayer. Micrographs in the bottom left corner have scale bars of $\SI{10}{\micro\meter}$. \textbf{(b,d)} Respective six-fold patterns reveal crystal orientation.}
\label{fig:Maps}
\end{figure}

Figure \ref{fig:Maps}a (\ref{fig:Maps}c) presents a spatial map of integrated SHG intensity revealing a prototypical MoSSe (WSSe) Janus monolayer flake with a micrograph in the bottom left (scale bar is $\SI{10}{\micro\meter}$). Figure \ref{fig:Maps}b (\ref{fig:Maps}d) outlines the MoSSe (WSSe) Janus monolayer crystal orientation extracted from polarization-dependent SHG measurement, revealing zig-zag and armchair crystal directions. 

%